\documentclass[letterpaper,twocolumn,showpacs,prl,aps]{revtex4-1}

\usepackage{amsmath}  
\usepackage{amsfonts} 
\usepackage{graphicx} 
\usepackage{amssymb}

\begin{document}

\title{Analog de Sitter space in a controlled Coulomb explosion}

\author{Eugene B. Kolomeisky}

\affiliation
{Department of Physics, University of Virginia, P. O. Box 400714, Charlottesville, Virginia 22904-4714, USA}

\date{\today}

\begin{abstract}
The Coulomb explosion of an atomic gas steered by laser-controlled time-dependent charging can be modeled by a macroscopic system of identical charges whose  number is not conserved.  We show that such a system can evolve in the spatially homogeneous and isotropic fashion that mimics accelerating cosmic expansions.  Specifically, for a constant rate of charge production, the resulting Friedmann-Coulomb equations have a stable fixed point corresponding to an analog of de Sitter space.  
\end{abstract}


\maketitle

The de Sitter (dS) space describes an exponentially expanding model of the world \cite{DeSitter} and plays a prominent role in cosmology.  In theories of the early Universe the dS space is central to resolving the puzzles of flatness, homogeneity, and isotropy of the Universe \cite{Guth,Starobinsky1} and explaining the formation of cosmic structures \cite{Starobinsky2}.   The dS metric is the homogenous and isotropic solution of Einstein's field equations in vacuum in the presence of the cosmological constant also known as the $\Lambda$-term.  The latter was introduced by Einstein \cite{Einstein} and has the effect of counteracting the long-range attractive gravitational interaction, making a static Universe possible.    Nowadays a consensus exists that the very early stage of rapid Universe expansion was due to the presence of matter having an unusual equation of state that can be imitated by a positive $\Lambda$ \cite{Mukhanov}.

After the discovery of the expansion of the modern Universe \cite{Friedmann,Le,Hubble}, the introduction of the $\Lambda$-term was regarded  as a complication that was not supported by observational data of the day \cite{LL2}.  Recently this attitude started to shift when an accelerating expansion of the Universe was established \cite{accelerated1,accelerated2}.  Presently a case can be made \cite{SS,CMB,bright} in favor of a positive $\Lambda$-term, also known as the dark energy \cite{dark}.

A dS space also appears in the steady-state model of the world \cite{steady-state} where the matter density in the expanding Universe remains fixed due to a continuous creation of matter.  We note that despite the lack of matter, the original dS space \cite{DeSitter} is also in a steady state.  Both versions of the dS space satisfy the perfect cosmological principle describing a world that is both space and time homogeneous and isotropic \cite{Mukhanov}.    

Despite its importance, experimentation with the dS space is impossible since the empirical foundation of cosmology is solely observational, and there is just one ongoing experiment.  The only possibility remaining (first pointed out by Unruh \cite{Unruh} in the context of the black hole physics) to test various cosmological scenarios is to look for analogous phenomena in areas of physics where laboratory experimentation can be carried out.  In condensed matter physics such possibilities are found with superfluid phases of $^{3}He$ \cite{Volovik}.  In atomic and molecular physics a similar role is played by Bose-Einstein-condensed gases \cite{Fedichev,BEC1,BEC2,BEC3,BEC4,BEC5,BEC6,BEC7}.  Specifically, there exists a proposal to realize a one-dimensional sonic dS space in an expanding cigar-shaped Bose-Einstein condensate \cite{Fedichev}.  While a full laboratory realization of these ideas is yet to come, a series of related phenomena -- specifically those mimicking the dynamical Casimir effect \cite{Casimir} and classical field dynamics in an expanding Universe \cite{pre} -- have already been observed.          

It was recently demonstrated \cite{EBK1} that the spatially homogeneous and isotropic evolutions of jellium - a one-component plasma of identical interacting charges in the presence of a uniform compensating charged background \cite{Wigner} - are governed by equations that have the structure of the Friedmann cosmological equations of the general theory of relativity \cite{Mukhanov}.  The resulting Hubble flows model oscillatory universes, including the anti-dS space.   Additionally, the dynamics of normal modes in the background of these homogeneous and isotropic solutions exhibits both red- and blue-shifting as well as the closely related effects of Hubble friction and anti-friction \cite{EBK2}.   

The jellium model finds applications in condensed matter physics \cite{Bohm_Pines,Pines_Nozieres,Mahan}, astrophysics \cite{Salpeter}, and in the limit of zero background charge it describes a Coulomb explosion \cite{CE}.  The latter can be caused by an interaction of an intense laser pulse with matter.  The large electric field of the pulse quickly drives the light electrons away from the heavy atoms, leaving positively charged ions behind; they undergo an explosion driven by the Coulomb repulsion.  Coulomb explosions mimic the non-singular accelerating open cosmologies in negatively curved spaces \cite{EBK1}.

The goal of this paper is to demonstrate that Coulomb explosions controlled by an external time-dependent charging can mimic a series of accelerating $(3+1)$-dimensional cosmologies including the dS space.  This can become a laboratory reality thanks to the modern laser technology that allows the active manipulation of the pulse-matter coupling \cite{control}, enabling the design of practically arbitrary optical waveforms including control of the phase, amplitude, and polarization \cite{Weiner}.  We note that the analogy between Coulomb explosion and uniform expansion of the early Universe has already been mentioned by Kaplan \textit{et al.} \cite{Kaplan} although conservative version of the effect considered by these authors does not mimic a dS space.

A controlled Coulomb explosion will  be described by a classical macroscopic theory with phenomenologically built-in charge creation.  Since the binding energy of the first electron is usually much smaller than the subsequent ones, there always exists a range of parameters of an ionizing source in which the system may be supposed to include only neutral atoms and singly-charged ions.  Hereafter it is assumed that the ionization process is nearly instantaneous, i.e. its time scale is much shorter than that of the ionic motion.   Additionally, if the system is a gas, we can neglect the motion of its neutral component  and only focus on the ions.    

We start by deriving the equations of hydrodynamics of an ideal liquid without matter conservation.  First, the continuity equation is modified by adding to its right-hand side a source term:
\begin{equation}
\label{modified_continuity}
\frac{\partial n}{\partial t}+\frac{\partial}{\partial x_{i}}n v_{i}=\Gamma
\end{equation} 
where  $n$ and $\textbf{v}$ are the number density and the velocity field, $\Gamma$ is the volume matter creation rate (i.e. the number of ions created per unit time per unit volume), and we used tensor notation.  The creation rate $\Gamma$ is uniform in space, as in most cases of interest the samples undergoing Coulomb explosion are small enough that the laser beam is an effectively uniform time-dependent electric field \cite{control}. The ions are created at the expense of neutral atoms whose density $N$ decreases as
\begin{equation}
\label{atom}
\frac{dN}{dt}=-\Gamma=-Nw(t)
\end{equation}
where $w(t)$ is the ionization rate.  Integrating Eq.(\ref{atom}) one finds
\begin{equation}
\label{reservoir}
N(t)=N_{0}\exp\left (-\int_{0}^{t}w(t')dt'\right )
\end{equation}
where $N_{0}$ is the initial density of neutral atoms.  The ionization rate $w(t)$ and thus the volume creation rate $\Gamma(t)=N(t)w(t)$ can be controlled via pulse shaping techniques \cite{control,Weiner}. 

Matter non-conservation also modifies the Euler equation.  Its form is obtained by starting with the momentum balance equation, i.e. Newton's second law applied to a small liquid volume       
\begin{equation}
\label{momentum_balance1}
\frac{\partial}{\partial t}n v_{i}=\frac{ne}{m} E_{i}-\frac{1}{m}\frac{\partial\Pi_{ik}}{\partial x_{k}},~~~\Pi_{ik}=p\delta_{ik}+mn v_{i}v_{k}
\end{equation}
where $e$ is the ion charge, $m$ is its mass, $\textbf{E}$ is the electric field, $\Pi_{ik}$ is the momentum flux density tensor and $p$ is the pressure \cite{LL6}.  It is assumed that the motion is non-relativistic and thus the magnetic effects are omitted.  Eq.(\ref{momentum_balance1}) can be transformed to 
\begin{eqnarray}
\label{momentum_balance2}
n\frac{\partial v_{i}}{\partial t}&+&v_{i}\left (\frac{\partial n}{\partial t}+\frac{\partial}{\partial x_{k}}n v_{k} \right )+n v_{k}\frac{\partial v_{i}}{\partial x_{k}}\nonumber\\
&=&\frac{ne}{m} E_{i}-\frac{1}{m}\frac{\partial p}{\partial x_{i}}.
\end{eqnarray}
Combining with Eq.(\ref{modified_continuity}) we arrive at the modified Euler equation
\begin{equation}
\label{modified_Euler}
\frac{\partial \textbf{v}}{\partial t} +(\textbf{v}\cdot \nabla)\textbf{v}= \frac{e}{m}\textbf{E}-\frac{\Gamma}{n}\textbf{v}-\frac{1}{mn}\nabla p.
\end{equation}
We see that matter non-conservation generates an effective friction force (per unit mass) $-\Gamma \textbf{v}/n$.  This effect has a singular character:  even if $\Gamma$ is small, the additional force might be large for $n$ small.  Since the ejected electrons are much lighter than the newly created ions,  the latter emerge with zero velocity relative to the neutral atoms; a force $-\Gamma \textbf{v}/n$ is required to accelerate them to the local flow velocity $\textbf{v}$.  While this makes the reference frame of the neutral atoms special, thus violating the frame independence of the physical laws, the modification (\ref{modified_Euler}) correctly describes the laser-excitation experiment.  This is then different from the Newtonian version of the steady-state theory \cite{Nsteady}, where it is assumed that the new particles are created with the local flow velocity; then the resulting Euler equation does not have an extra friction force.  Although the latter model preserves the frame independence of the physical laws, it does not seem to be experimentally realizable.    

Finally, the electric field $\textbf{E}$ is determined by Gauss's law
\begin{equation}
\label{electric_Gauss}
\nabla\cdot\textbf{E}=4\pi en.
\end{equation}

Following Ref. \cite{EBK1} we seek solutions to Eqs.(\ref{modified_continuity}), (\ref{modified_Euler}), and (\ref{electric_Gauss}) corresponding to a spatially homogeneous and isotropic ion density $n=n(t)$.  Then the pressure $p$ (which is a function of $n$) is also uniform, so that $\nabla p=0$.  As a result, the last term in the right-hand side of the Euler equation (\ref{modified_Euler}) is identically zero.  The space isotropy also dictates that in the rest frame of one of the ions the remainder of the ionic liquid is characterized by a radially symmetric velocity field, $\textbf{v}(\textbf{r},t)=v(r,t)\textbf{r}/r$, where the radius vector $\textbf{r}$ is the position relative to the ion at rest.  Then the dynamics of the ionic liquid is governed by the electric field that follows from Gauss's law (\ref{electric_Gauss}):          
\begin{equation}
\label{electric_field}
\textbf{E}=\frac{4\pi e}{3}n(t)\textbf{r}.
\end{equation}
With all this in mind, the rate (\ref{modified_continuity}) and the Euler (\ref{modified_Euler}) equations can be written as
\begin{equation}
\label{radial_continuity}
\dot{n}+n\frac{(r^{2}v)'}{r^{2}}=\Gamma(t)
\end{equation} 
\begin{equation}
\label{radial_Euler}
\frac{\dot{v}+vv'}{r}=\frac{4\pi e^{2}}{3m}n-\frac{\Gamma(t)}{n}\frac{v}{r}
\end{equation}
where the dot and the prime are shorthands for the derivatives with respect to $t$ and $r$.  The second term in the left-hand side of the rate equation (\ref{radial_continuity}) is strictly time-dependent if the ratio $(r^{2}v)'/r^{2}$ is a function of time:
\begin{equation}
\label{Hubble}
\frac{(r^{2}v)'}{r^{2}}=3H(t)
\end{equation}
A solution to Eq.(\ref{Hubble}) in the form of Hubble's law 
\begin{equation}
\label{Hubble_law}
\textbf{v}=H(t)\textbf{r}
\end{equation}
with $H(t)$ being the Hubble parameter \cite{Mukhanov} transforms the rate (\ref{radial_continuity}) and Euler (\ref{radial_Euler}) equations into a system of two first-order differential equations for the two unknown functions $n(t)$ and $H(t)$: 
\begin{equation}
\label{Hubble_definition}
\dot{n}=\Gamma(t)-3Hn
\end{equation}
\begin{equation}
\label{Friedmann1}
\dot{H}+H^{2}(t)=\frac{4\pi e^{2}}{3m}n-\frac{\Gamma(t)}{n}H
\end{equation}
The conservative $\Gamma=0$ limit of this system of equations was studied previously \cite{EBK1} where its mapping to the Friedmann cosmological equations \cite{Mukhanov} was established.  To emphasize the underlying Coulomb nature of the problem, these will be referred to as the Friedmann-Coulomb (FC) equations in what follows.

Since $\Gamma(t)>0$, the creation of charge (represented by the first term in the right-hand side of Eq.(\ref{Hubble_definition})) counteracts the decrease of the density due to the expansion, $H(t)>0$ (represented by the second term).  According to Eq.(\ref{Friedmann1}), the charging also opposes the growth of the Hubble parameter.   

\begin{figure}
\includegraphics[width=1.0\columnwidth, keepaspectratio]{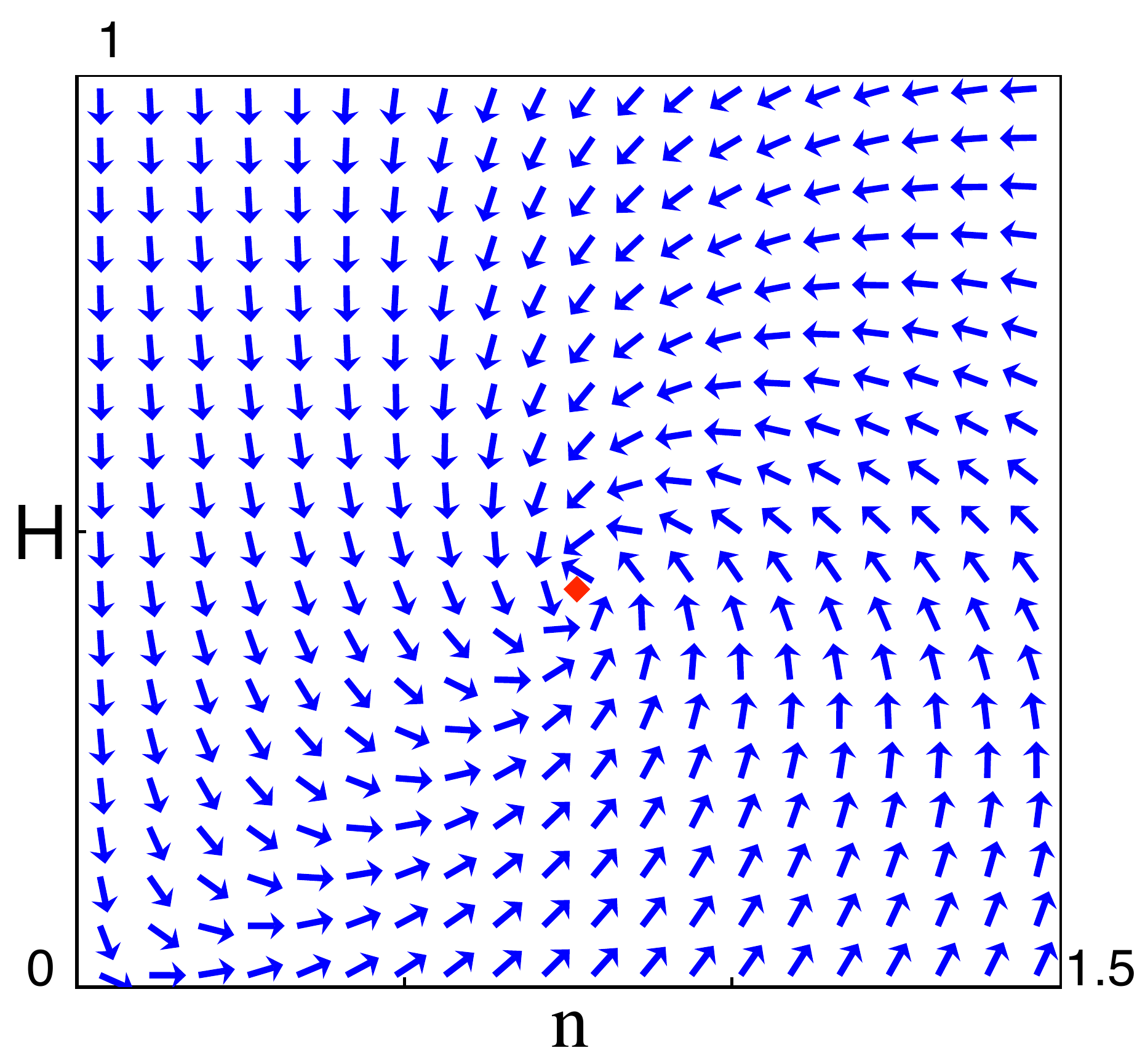} 
\caption{(Color online) Phase portrait of the Friedmann-Coulomb equations (\ref{FC1}) and (\ref{FC2}) in units defined in Eq.(\ref{units}).  The blue arrows indicate the directions of evolution of the density $n$ and the Hubble parameter $H$ with time.  The stable de Sitter fixed point $n^{*}=4/12^{2/3}$ and $H^{*}=1/12^{1/3}$ is shown as a red diamond.}
\label{portrait}
\end{figure}

Since the volume rate $\Gamma(t)$ is an externally controllable parameter, a series of accelerating cosmic expansions can be mimicked in a controlled Coulomb explosion.  Below we limit ourselves to the case of a constant rate of charging.  For $\Gamma=const$ it is convenient to introduce the following units of time and density, respectively
\begin{equation}
\label{units}
[t]=\left (\frac{3m}{4\pi e^{2}\Gamma}\right )^{1/3}, ~~~[n]=\left (\frac{3m\Gamma^{2}}{4\pi e^{2}}\right )^{1/3}
\end{equation}
With this choice the FC equations (\ref{Hubble_definition}) and (\ref{Friedmann1}) acquire the parameter-free form
\begin{equation}
\label{FC1}
\dot{n}=1-3Hn
\end{equation}
\begin{equation}
\label{FC2}
\dot{H}=n-H^{2}-\frac{H}{n}
\end{equation}
The phase portrait of the FC equations (\ref{FC1}) and (\ref{FC2}) is shown in Figure \ref{portrait}.  
The primary consequence of the charge creation is the existence of a stable fixed point $n^{*}=4/12^{2/3}$ and $H^{*}=1/12^{1/3}$ which means that for any initial conditions the system asymptotically approaches the regime of constant density $n^{*}$ and expansion rate $H^{*}$.  The latter is the hallmark of the dS space.  Indeed the Hubble law (\ref{Hubble_law}) can be alternatively stated in terms of a new function $a(t)$ called the scale factor \cite{Mukhanov} such as 
\begin{equation}
\label{scale_factor}
H(t)=\frac{\dot{a}}{a}.
\end{equation}
The function $a(t)$ describes expansion of the system:  substituting this expression into Hubble's law (\ref{Hubble_law}), employing $\textbf{v}=d\textbf{r}/dt$, and integrating following the motion we find
\begin{equation}
\label{comoving}
\textbf{r}=a(t)\textbf{x}
\end{equation}
where $\textbf{x}$ is a time-independent co-moving coordinate vector defining the particle considered.  The vector $\textbf{x}$ factors out the dynamics and the scale factor $a(t)$ relates it to the proper (laboratory) position of the particle $\textbf{r}$.   If $H(t\rightarrow \infty)\rightarrow H^{*}=const$, asymptotically the scale factor increases exponentially, $a(t\rightarrow \infty)\propto \exp(H^{*}t)$.  It is curious that the dS fixed point $(n^{*}, H^{*})$ is approached in an oscillatory manner.  

The dS space realized in a controlled Coulomb explosion is a relative of that of the steady-state models of the world \cite{steady-state}:  the amount of created charge exactly compensates for the decrease of the density due to the expansion.  

From the experimental standpoint it is important to know how sensitive are our conclusions to perturbations away from uniformity.  For the conservative $\Gamma=0$ case this problem was already addressed \cite{EBK2} with the conclusion that uniform Coulomb explosion is stable with respect to small perturbations.  The same is expected to hold in the presence of charging due to faster expansion which is even more efficient in stretching of perturbations than in the conservative case.    

Finally, we have to discuss what it takes to sustain $\Gamma=const$ and reach the asymptotic dS limit in the laser-excitation experiment.  First, we observe that the dS limit will be effectively reached in a time which in the original physical units is given by the scale $[t]$ in Eq.(\ref{units});  this is the lower bound on the duration of the experiment.  At the same time if $\Gamma=const$, according to Eq.(\ref{atom}) the density of neutral atoms decreases as $N(t)=N_{0}-\Gamma t$ which means that in a time $N_{0}/\Gamma$ all the atoms will be ionized.  Requiring that $[t]\ll N_{0}/\Gamma$ or equivalently 
\begin{equation}
\label{condition}
\frac{m\Gamma^{2}}{e^{2}N_{0}^{3}}\ll 1
\end{equation}
will guarantee both the constant volume rate of charging and achievement of the dS limit of Coulomb explosion during the experiment.  The condition (\ref{condition}) can be always satisfied by choosing sufficiently small ion production rate $\Gamma$.  Simultaneously small $\Gamma$ increases the lower bound $[t]$ (\ref{units}) on the duration of the experiment.  

In the past two specific experimental setups have been successful in controlling Coulomb explosions:  pulse duration and laser intensity variations allowed for experimental control of the explosion of $H_{2}$ \cite{Hydrogen}, and a dual-pulse excitation scheme made it possible to control the explosion of Silver clusters \cite{dual}.  These techniques also hold a promise to trigger the Coulomb explosion that mimics the dS space.

To summarize, we demonstrated how a series of analog accelerating cosmological models can be realized in controlled Coulomb explosion experiments.  Specifically, the dS space is an inevitable outcome for the constant volume charging rate.  The experimental signature of the dS regime is an expansion where both the density and the Hubble parameter are constant;  these fixed values are approached in an oscillatory manner.  We hope that our predictions can be experimentally tested in the near future.  While existing proposals of analog cosmological evolutions require careful tuning (see, for example, Ref. \cite{Fedichev}) ours does not.              

I am grateful to J. P. Straley for a discussion.

\end{document}